\documentclass[a4paper]{article}

\usepackage[margin=1in]{geometry} 

\usepackage{amsmath}
\usepackage{amsthm}
\usepackage{amssymb}
\usepackage{physics}

\usepackage[utf8]{inputenc}
\usepackage{hyperref}
\hypersetup{
	unicode,
	pdfauthor={Author One, Author Two, Author Three},
	pdftitle={A simple article template},
	pdfsubject={A simple article template},
	pdfkeywords={article, template, simple},
	pdfproducer={LaTeX},
	pdfcreator={pdflatex}
}

\usepackage[super,sort&compress,numbers]{natbib}

\usepackage{graphicx, color}
\usepackage[labelfont=bf,labelsep=space]{caption}
\usepackage{algorithm, algpseudocode} 
\usepackage{mathrsfs} 

\newcommand{\timespan}{912}
\newcommand{\uptime}{591}

\newcommand{\cmunu}{c_{\mu\nu}}
\newcommand{\cmn}{c_{MN}}
\newcommand{\Clv}{C_0^{(2)}}
\newcommand{\kquad}{\kappa_{\mathrm{q}}}
\newcommand{\klv}{\kappa_{\mathrm{LV}}}
\newcommand{\kappaVal}{0.13(3)}
\newcommand{\darktime}{1.15}
\newcommand{\rfFreqDrift}{1.6}
\newcommand{\divqpn}{29(3)}
\newcommand{\tbunch}{15}

\newcommand{\chsqfit}{0.92}
\newcommand{\omegaSD}{\omega_{\oplus}}
\newcommand{\pf}{P_{\mathrm{f}}}

\newcommand*{\citen}[1]{%
	\begingroup
	\romannumeral-`\x 
	\setcitestyle{numbers}%
	\cite{#1}%
	\endgroup   
}

\title{Improved bounds on Lorentz violation from composite-pulse Ramsey spectroscopy in a trapped ion}
\author{Laura S. Dreissen$^1$\thanks{Corresponding author. Email: laura.dreissen@ptb.de}, Chih-Han Yeh$^1$, Henning A. F\"{u}rst$^{1,2}$, \\Kai C. Grensemann$^1$, Tanja E. Mehlst\"{a}ubler$^{1,2}$}

\date{
	$^1$Physikalisch-Technische Bundesanstalt, Bundesallee 100, 38116, Braunschweig, Germany \\ %
	$^2$Institut f\"{u}r Quantenoptik, Leibniz Universit\"{a}t Hannover, Welfengarten 1, 30167, Hanover, Germany \\[2ex]%
}

\begin{document}
	\maketitle
	
	\begin{abstract}
		\textbf{In attempts to unify the four known fundamental forces in a single quantum-consistent theory, it is suggested that Lorentz symmetry may be broken at the Planck scale. Here we search for Lorentz violation at the low-energy limit by comparing orthogonally oriented atomic orbitals in a Michelson-Morley-type experiment. We apply a robust radiofrequency composite pulse sequence in the $^2F_{7/2}$ manifold of an Yb$^+$ ion, extending the coherence time from 200\,$\mu$s to more than 1\,s. In this manner, we fully exploit the high intrinsic susceptibility of the $^2F_{7/2}$ state and take advantage of its exceptionally long lifetime. We match the stability of the previous best Lorentz symmetry test nearly an order of magnitude faster and improve the constraints on the symmetry breaking coefficients to the 10$^{-21}$ level. These results represent the most stringent test of this type of Lorentz violation. The demonstrated method can be further extended to ion Coulomb crystals.}
	\end{abstract}
	
	\section{Introduction}
	
	The standard model (SM) of particle physics describes non-gravitational interactions between all particles and fields, while gravitation is described by general relativity in a classical manner. Together, they have explained many physical phenomena observed in the universe remarkably well, but an accurate description of gravity at the quantum level is lacking. A number of theories that attempt to unify the SM and gravitation at the Planck scale suggest that breaking of Lorentz symmetry might occur either spontaneously \cite{kostelecky_spontaneous_1989} or explicitly  \cite{hoifmode_quantum_2009,pospelov_Lorentz_2012,cognola_covariant_2016}. Lorentz symmetry states that the outcome of a local experiment does not depend on the orientation or the velocity of the apparatus \cite{tobar_testing_2010}. A suppressed effect emerging from Lorentz violation (LV) at the Planck scale could be observed at experimentally accessible energies in the laboratory \cite{mattingly_modern_2005}. Accurate spectroscopic measurements in trapped particles have reached fractional uncertainties beyond the natural suppression factor \cite{kostelecky_cpt_1995}, which makes a hypothetical LV measurable in such systems. Furthermore, at high energies, LV could be suppressed by super-symmetry \cite{nibbelink_lorentz_2005}. Therefore, accurate low-energy measurements in atoms are suitable to search for LV and complement existing bounds set at high energies with, e.g.,~particle colliders and astrophysical observations \cite{abazov_search_2012,aaij_search_2016,schreck_vacuum_2017,albert_constraints_2020}.\\ 
	Laboratory tests of Lorentz symmetry are based on a similar principle as introduced by Michelson and Morley, who used a rotating interferometer to measure the isotropy of the speed of light \cite{michelson_influence_1887}. Improved bounds on LV for photons have been realized by a variety of experiments involving high-finesse optical and microwave cavities, see e.g. refs.~\citen{herrmann_rotating_2009,muller_modern_2003,tobar_testing_2010,eisele_laboratory_2009,safronova_search_2018}. Spectroscopic bounds for protons and neutrons are set using atomic fountain clocks \cite{wolf_cold_2006,pihan-le_Lorentz-symmetry_2017} and co-magnetometers \cite{brown_new_2010,smiciklas_new_2011}. More recently, the bounds on LV in the combined electron-photon sector have been explored using precision spectroscopy in trapped ions \cite{pruttivarasin_michelson-morley_2015,megidish_improved_2019,sanner_optical_2019}. These experiments compare energy levels with differently oriented, relativistic, non-spherical electron orbitals as the Earth rotates. Strong bounds on LV were set with trapped $^{40}$Ca$^{+}$ ions, where a decoherence-free entangled state of two ions in the electronic $^2D_{5/2}$ manifold was created to suppress ambient noise \cite{pruttivarasin_michelson-morley_2015,megidish_improved_2019}. The relatively short $1.2$\,s radiative lifetime of the $^2D_{5/2}$ state in Ca$^+$ and the requirement for high-fidelity quantum gates limit the scalability and, ultimately, the sensitivity of LV tests with this scheme \cite{megidish_improved_2019}. The highly relativistic $^2F_{7/2}$ state in the Yb$^{+}$ ion is an order of magnitude more sensitive to LV than the $^2D_{5/2}$ state in Ca$^{+}$ \cite{dzuba_strongly_2016,shaniv_new_2018} and its radiative lifetime was measured to be about $1.6$ years \cite{lange_lifetime_2021}. However, gate operation on the electronic octupole (E3) transition, required to efficiently populate the $^2F_{7/2}$ manifold, suffers from a low fidelity, making entanglement-based techniques unfeasible. The beneficial properties of Yb$^+$ were recently partly exploited in a 45-day comparison of two separate state-of-the-art single-ion optical $^{171}$Yb$^{+}$ clocks, both with a fractional uncertainty at the $10^{-18}$ level, reaching a more than ten-fold improvement \cite{sanner_optical_2019}. However, operating the Yb$^+$ ion as an optical clock limits the experiment to only probe the Zeeman levels that are least sensitive to LV ($m_j=\pm1/2$).\\
	In this work we present improved bounds on LV in the electron-photon sector using a method that fully exploits the high susceptibility of the stretched $m_j=\pm7/2$ states in the $^2F_{7/2}$ manifold of Yb$^+$ and takes advantage of its long radiative lifetime. With a robust radiofrequency (rf) spin-echoed Ramsey sequence \cite{shaniv_new_2018,genov_arbitrarily_2017} we populate all Zeeman sublevels in the $^2F_{7/2}$ manifold and make a direct energy comparison between the orthogonally oriented atomic orbitals within a single trapped ion. We decouple the energy levels from noise in the ambient magnetic field to reach a 5000-fold longer coherence time during the Ramsey measurement and extend the dark time to $T_{\mathrm{D}}>1$\,s. With an unprecedented sensitivity to LV, scaling as $\sigma_{\mathrm{LV}} \propto \sqrt{1/T_{\mathrm{D}}N_{\mathrm{ion}}}$, we reduce the required total averaging time by nearly an order of magnitude already with a single ion ($N_{\mathrm{ion}}=1$). The measurement scheme is simple, robust and scalable and eliminates the requirement of optical clock operation or high-fidelity quantum gates. The rf sequence is insensitive to both temporal and spatial field inhomogeneities and can be applied to a string of $N_{\mathrm{ion}}$ trapped ions for an increased sensitivity to LV in the future.
	
	\section{Results}
	\subsection{Theoretical framework}
\begin{figure}[h]
		\centering	
		\includegraphics[width=0.5\textwidth]{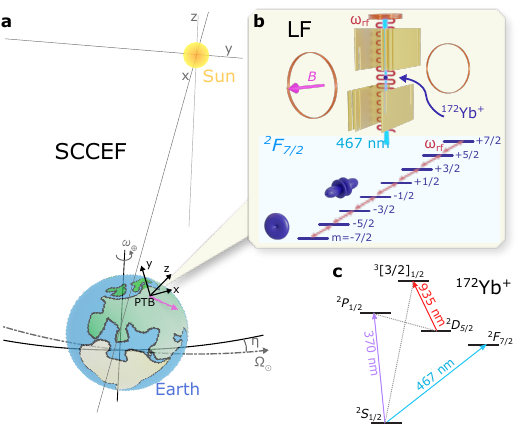}
		\caption{A schematic overview of the experimental principle. \textbf{a} In the laboratory frame (LF) coordinate system $z$ points vertically upwards, $y$ points North and $x$ points east. The fixed quantization axis $\mathbf{B}$ (pink arrow) in the LF coordinate system lies in the horizontal plane and points $20^{\circ}$ south of $x$ and probes different directions in the SCCEF as the Earth rotates around its axis ($\omegaSD$) and orbits the Sun ($\Omega_{\odot}$). \textbf{b} In the LF a single $^{172}$Yb$^+$ ion is trapped in a segmented rf trap. Two coils generate a quantization field of $221$\,$\mu$T. The electron orbitals of the $m_j=\pm1/2$ and $m_j=\pm7/2$ Zeeman sublevels in the $^2F_{7/2}$ state orient themselves orthogonally to each other. The $^2F_{7/2}$ state is population via the E3 transition near $467$\,nm. The $m_j$ substates are coupled via the rf magnetic field at $\omega_{\mathrm{rf}}$ created with a coil that is placed $5.5$\,cm above the ion. \textbf{c} A simplified energy level diagram of $^{172}$Yb$^+$, showing the optical E3 transition and the transitions near $370$\,nm and $935$\,nm used for Doppler cooling and state detection.}
		\label{fig:exp_scheme}
	\end{figure}

	The constraints on LV extracted in this work are quantified in the theoretical framework of the standard model extension (SME) \cite{lorentz_colladay_1998}. The SME is an effective field theory in which the SM Lagrangian is extended with all possible terms that are not Lorentz invariant. It is a platform in which LV of all SM particles are described, enabling comparisons between experimental results from many different fields \cite{kostelecky_data_2011}. In spectroscopic experiments, a violation of Lorentz symmetry can be interpreted as LV of electrons or photons, because there is no preferred reference system. In this work, we interpret the results as a difference in isotropy between photons and electrons, similar as in refs.~\citen{pruttivarasin_michelson-morley_2015,sanner_optical_2019}.\\
	LV in the combined electron-photon sector is quantified by adding a symmetry-breaking tensor $\cmunu'=\cmunu+k_{\mu\nu}/2$ to the SM Lagrangian \cite{lorentz_colladay_1998,dzuba_strongly_2016}, where $\cmunu$ and $k_{\mu\nu}$ describe LV for electrons and photons, respectively. For simplicity, the prime is omitted throughout the rest of this work and the extracted coefficients are those of the combined $\cmunu'$ tensor, which is taken as traceless and symmetric. The components of the $\cmunu$ tensor are frame dependent. A unique definition of the symmetry breaking tensor $\cmn$ exists in the Sun-centered, celestial, equatorial frame (SCCEF), illustrated in Fig.\,\ref{fig:exp_scheme}\,a. In order to make comparisons with other experiments, the $\cmunu$ tensor defined in our local laboratory frame is transformed to the SCCEF to constrain the components of the $\cmn$ tensor. The full derivation of the transformation can be found in the Supplementary Information. \\ 
	In a bound electronic systems, LV leads to a small energy shift \cite{dzuba_strongly_2016,shaniv_new_2018}
	\begin{equation}
		\label{eq:delta_ham}
		\delta \mathcal{H} = -\frac{1}{6m_e}\Clv T_0^{(2)}\,,
	\end{equation}
	where $m_e$ is the electron mass, the $T_0^{(2)}=\mathbf{p}^2-3p_z^2$ operator depends on the direction of the electron's momentum and $\Clv $ contains elements of the $\cmn$ tensor. For a state with total angular momentum $J$ and projection $m_j$ onto the quantization axis $\hat{z}$, the matrix element of the $T_0^{(2)}$ operator is given by \cite{shaniv_new_2018}
	\begin{equation}
		\label{eq:Tmatrixel}
		\bra{J,m_j}T_0^{(2)}\ket{J,m_j}=\frac{-J(J+1)+3m_j^2}{\sqrt{(2J+3)(J+1)(2J+1)J(2J-1)}}\times \langle J\| T^{(2)}\|J \rangle\,.
	\end{equation}
	Equations~(\ref{eq:delta_ham}) and (\ref{eq:Tmatrixel}) show that in the SCCEF, LV manifests itself as an energy shift that modulates with Earth's rotation frequency. The magnitude of this shift is dependent on both $m_j^2$ and the reduced matrix element $\langle J\| T^{(2)}\|J \rangle$. The value of the latter is particularly high for the $^2F_{7/2}$ manifold in the Yb$^+$ ion \cite{dzuba_strongly_2016,shaniv_new_2018}. The goal of this experiment is to test LV in a single trapped $^{172}$Yb$^+$ ion by measuring the energy difference between $m_j$ substates in the $^2F_{7/2}$ manifold as the Earth rotates.
	\subsection{Measurement principle}
\begin{figure}[h]
		\centering	
		\includegraphics[width=0.5\textwidth]{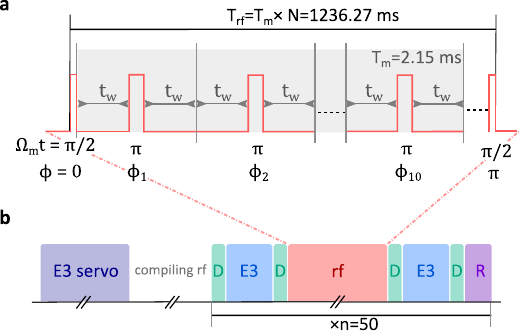}
		\caption{Pulse sequences. \textbf{a} The applied composite rf pulse sequence is a more robust version of a spin-echoed Ramsey scheme. A Ramsey sequence of $T_{\mathrm{rf}}=1236.27$\,ms starts and ends with a $\pi/2$-pulse. A $2.15$\,ms-long modulation sequence of 10 $\pi$-pulses with phases $\phi_i$ ($i=1...10$), each spaced by $2t_{\mathrm{w}}=200$\,$\mu$s, is repeated $N=575$ times to cancel dephasing due to ambient magnetic field noise. The phases $\phi_i$ are chosen such (see text) that the sequence is robust against pulse errors. \textbf{b} Every $n=50$ measurements, a 4-point servo sequence (E3 servo) of $40$\,s is applied to follow the E3 center frequency, after which the rf sequence is compiled for $2.5$\,s. The measurement sequence consists of two $50$\,ms-long E3 pulses (E3), the rf pulse sequence (rf) of $1236.27$\,ms, four detection pulses (D) of $2.5$\,ms and a sequence at the end of $20$\,ms for state preparation (R) for the next measurement run. During post-selection, the data points are considered valid if the ion was in the desired state during the first three detection pulses.}
		\label{fig:pulse_sequences}
	\end{figure}

	The experiment is performed with a single ion, stored in a linear rf Paul-trap, see Fig.\,\ref{fig:exp_scheme}\,b. It is cooled to the Doppler limit of around $0.5$\,mK on the dipole allowed transition near a wavelength of $370$\,nm, assisted by a repumper near $935$\,nm. A set of coils is used to define the quantization axis of $B=221$\,$\mu$T, which lies in the horizontal plane with respect to Earth's surface and points 20$^{\circ}$ south of east, see Fig.\,\ref{fig:exp_scheme}\,a. Active feedback is applied on auxiliary coils in three dimensions to stabilize the magnetic field. The $^2F_{7/2}$ state can be efficiently populated via coherent excitation of the highly-forbidden electric octupole (E3) transition (Fig.\,\ref{fig:exp_scheme}\,c) using an ultra-stable frequency-doubled laser at $934$\,nm \cite{fuerst_coherent_2020}, which is either stabilized only to a cryogenic silicon cavity via a frequency comb \cite{matei_1.5micron_2017} or, optionally, to the single-ion optical $^{171}$Yb$^+$ clock \cite{huntemann_single-ion_2016}. A Rabi frequency of $\Omega_{\mathrm{E3}}/2\pi=10$\,Hz is achieved on the E3 transition. More details on the experimental apparatus can be found in the Methods section. \\
	The free evolution of a substate $\ket{^{2}F_{7/2},m_j}$ interacting with a magnetic field $\mathbf{B}\equiv B_z \hat{z}$, is given by the Hamiltonian $\mathcal{H}_{\mathrm{free}}=\mathcal{H}_{\mathrm{lin}}+\mathcal{H}_{\mathrm{quad}}=\mu B_z J_z+\kappa J_z^2$, where $\mu$ is the magnetic moment. The quadratic term in the Hamiltonian gives rise to an energy shift according to $E_{m_j}/h \equiv \kappa/2 \pi \times m_j^2$. The value of $\kappa=\kquad+\klv$ is dependent on the quadrupole shift for the trapped ion in the $^2F_{7/2}$ state and a possible shift due to LV, respectively. The contribution from LV to $\kappa$ is given by \cite{shaniv_new_2018}
	\begin{equation}
		\label{eq:kappa_sens_lv}
		\frac{\klv}{2\pi}=5.1\times10^{15}\,\mathrm{Hz}\times C_0^{(2)}\,,
	\end{equation}
	where $C_{0}^{(2)}$ contains components of the $\cmn$ tensor in the SCCEF, see the Supplementary Information.\\
	A modulation of the quadratic contribution to the Zeeman splitting in the $^2F_{7/2}$ manifold is measured with rf spectroscopy. The $m_j$ levels are coupled via a rf magnetic field supplied to the ion by a resonant LC circuit. The coupling term in the Hamiltonian is given by $\mathcal{H}_{\mathrm{coupling}}=\Omega_{F} \cos(\omega_{\mathrm{rf}} t+\phi)J_{x}$, where $\Omega_{F}/2 \pi= 33$\,kHz is the multilevel Rabi frequency and $\omega_{\mathrm{rf}}$ and $\phi$ are the frequency and the phase of the rf field, respectively. The rf frequency is close to resonance with the, to first order, equidistant $m_j$ levels given by $\omega_{\mathrm{rf}}/2\pi=\mu B_z/h+\delta(t)/2\pi \approx3.5$\,MHz, where $\delta(t)$ is a small detuning from temporal drifts in the ambient magnetic field. The full Hamiltonian of the system in the interaction picture after applying the rotating wave approximation is given by
	\begin{equation}
		\label{eq:CouplingHamrf}
		\mathcal{H}=\delta(t)J_z+\kappa J_z^2+\Omega_{F}[J_x \cos(\phi)-J_y \sin(\phi)]\,,
	\end{equation}
	where the changes in $\delta(t)$ and $\kappa$ should be much slower than the $\pi$-pulse time of $t_{\pi}=\pi/\Omega_{F}=15$\,$\mu$s.\\
	A composite rf pulse sequence, based on a spin-echoed Ramsey scheme, is implemented to mitigate the influence of $\delta(t)$, while retaining a high sensitivity to variations of $\kappa$. A schematic overview of the rf sequence is shown in Fig.\,\ref{fig:pulse_sequences}\,a. Starting from either one of the $m_j=\pm1/2$ states, a $\pi/2$-pulse, i.e.~$t_{\pi/2}=t_{\pi}/2$, with phase $\phi=0$ spreads the population over all the Zeeman sublevels. A modulation sequence, consisting of ten repetitions of  the form $[t_{\mathrm{w}}]-[t_{\pi}(\phi_i)]-[t_{\mathrm{w}}]$, is applied to cancel dephasing from $\delta(t)$. Here $t_{\mathrm{w}}$ is the time in which the state freely evolves and $t_{\pi}(\phi_i)$ indicates a $\pi$-pulse with phase $\phi_{i}$. In our experimental environment, dephasing is successfully canceled for $t_{\mathrm{w}}=100$\,$\mu$s. The phases $\phi_{i}$ of the consecutive $\pi$-pulses are set to $\left(0,4,2,4,0,0,4,2,4,0\right)\pi/5$, for which the modulation sequence is shown to be highly robust against pulse errors from, e.g., detuning and intensity variations \cite{genov_arbitrarily_2017}. At the end of the sequence, a $\pi/2$-pulse with phase $\phi=\pi$ retrieves a fraction $\pf$ of the population back into the initial $m_j=\pm1/2$ state. The Ramsey dark-time, i.e.~the time in which the state freely evolves, can be extended to $T_{\mathrm{D}}=N\times20\,t_{\mathrm{w}}$ by repeating the modulation sequence $N$ times.\\ 
	The retrieved fraction $\pf$ is dependent on the phase $\kappa T_{\mathrm{D}}$ acquired during the free-evolution time, which stems from the quadratic term in the Hamiltonian of equation (\ref{eq:CouplingHamrf}). Therefore, $\klv$ can be extracted from a measurement of $\pf$ at a fixed $T_{\mathrm{D}}$. Since a hypothetical LV would manifest itself as a modulation of $\kappa$ at frequencies related to a sidereal day ($\omegaSD/2\pi=1/23.934$\,h), a signal of LV is characterized by oscillations of $\pf$ at the same modulation frequencies. A measurement of $\pf$ is realized by de-exciting the ion via the E3 transition to the $^2S_{1/2}$ state, where it can be detected by collecting fluorescence on the dipole allowed $^2S_{1/2}\rightarrow^2P_{1/2}$ transition.\\
	The highest measurement accuracy is achieved when the measured quantity $\pf$ is most sensitive to variations of $\kappa$, i.e.~$\vert\mathrm{d} \pf/\mathrm{d}\kappa\vert$, is maximized, which is the case at $\kappa T_{\mathrm{D}}=0.15$\,rad \cite{shaniv_new_2018}. Using the axial secular frequency of $266(5)$\,kHz set during the measurement campaign and the quadrupole moment of the $^2F_{7/2}$ state \cite{lange_coherent_2020}, $\kquad$ was calculated to be $\kappaVal$ rad/s, see the Supplementary Information. The corresponding optimal Ramsey dark time is $T_{\mathrm{D}}=\darktime$\,s.\\
	
	\subsection{Experimental sequence}
	A schematic overview of the full experimental sequence is shown in Fig.\,\ref{fig:pulse_sequences}\,b. Stable long-term operation of the experiment is required to resolve oscillation periods related to a potential LV, i.e.~11.967 and 23.934 hours. Especially the center frequency of the E3 transition is sensitive to external perturbations from, e.g.,~magnetic field drifts and intensity fluctuations. Therefore, a 4-point servo-sequence of two opposite Zeeman transitions $\ket{^2S_{1/2},m_j=\pm1/2}\rightarrow\ket{^2F_{7/2},m_j=\pm1/2}$ at half the linewidth is applied every $n=50$ measurement runs to follow the E3 center frequency. For details on this technique, see e.g. ref~\citen{ludlow_optical_2015}. On average a population transfer of $80\,\%$ is realized to the $^2F_{7/2}$ state via the E3 transition using the servo-sequence.\\   
	After the E3 servo and $2.5$\,s of compilation time for the rf pulse sequence, a measurement starts with a detection pulse to determine if the ion is cooled and prepared in the correct initial $\ket{^2S_{1/2},m_j=\pm1/2}$ state. If this is not the case, consecutive repumping and recooling sequences are applied. To save overhead time, either the $\ket{^2F_{7/2},m_j=+1/2}$ or the $\ket{^2F_{7/2},m_j=-1/2}$ state is populated via E3 excitation of the same Zeeman transition as was addressed last by the servo sequence. Another detection pulse is applied to determine if the E3 excitation was successful. A single rf modulation sequence of $T_{\mathrm{m}}=10\times t_{\pi}+20\times t_{\mathrm{w}}=2.15$\,ms is repeated $N=575$ times to reach $T_{\mathrm{D}}=1.15$\,s. The full rf sequence runs for $T_{\mathrm{rf}}=N\times T_{\mathrm{m}}+2\times t_{\pi/2}=1236.27$\,ms. A third detection pulse is applied to determine if the ion was quenched to the ground state during the rf sequence due to, e.g.,~a collision with background gas. The ion is de-excited on the E3 transition and, with a final detection pulse, $\pf$ is measured. At the end of the measurement sequence, the ion is re-cooled and prepared in the required $\ket{^2S_{1/2},m_j=\pm1/2}$ electronic ground state for the next measurement run. Data points are only considered valid when the ion was in the correct state at both the second and the third detection stage. After post-selection of valid data points an average contrast of $0.77(6)$ is reached. Including additional overhead from compilation time, data points are obtained at a rate of 1/191\,s$^{-1}$.  
	
	\subsection{Bounds on Lorentz violation}
\begin{figure}[h]
		\centering	
		\includegraphics[width=1\textwidth]{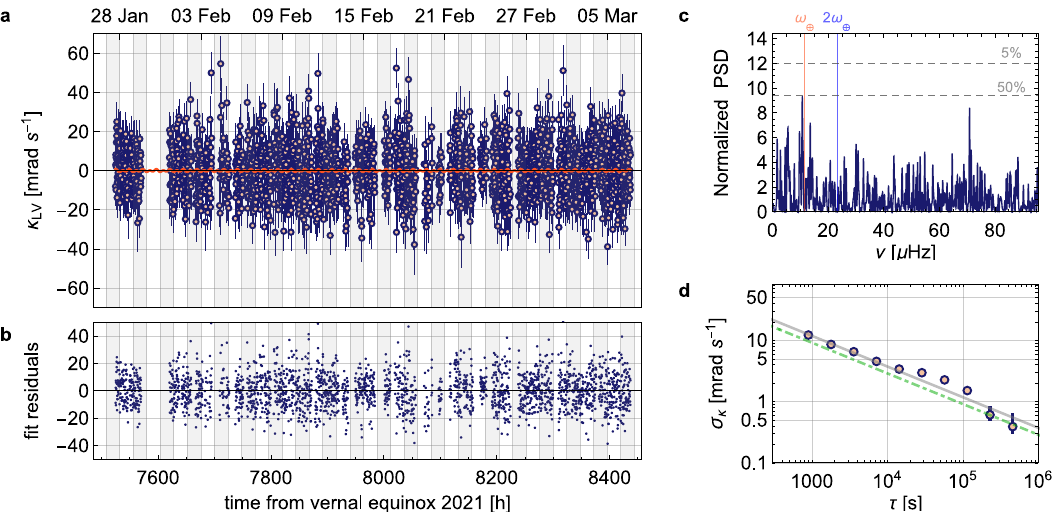}
		\caption{Data acquired over a period of \timespan\,h with an up-time of \uptime\,h. \textbf{a} The measured value of $\klv$ (blue data points) extracted from the retrieved population $\pf$. Data points are binned at $15$ minute intervals. The error bars reflect the standard error ($1\sigma$) from quantum projection noise. The data is fitted to equation~(\ref{eq:fit_func}) (red line) from which bounds are set on components of the $\cmn$ tensor. \textbf{b} The residuals of the fit with a reduced $\chi^2=\chsqfit$. \textbf{c} A Lomb-Scargl periodogram for spectral analysis of irregularly spaced data. The normalized power spectral density is plotted for frequencies between 0 and 100\,$\mu$Hz. The vertical lines indicate the sidereal day frequency, $\omegaSD$, (red) and its second harmonic, $2\omegaSD$, (blue). The dashed lines indicate the statistical significance level, i.e.~p-value, of the peaks \cite{glynn_detecting_2005}. \textbf{d} The Allan deviation, $\sigma_{\kappa}$, of the data points shown in \textbf{a} as a function of averaging time, $\tau$. The error bars correspond to the standard error of the stability. From a fit (gray line) to the data, the stability is extracted to be $\sigma_{\kappa}=372(9)\,\mathrm{mrad\,s^{-1}}\,\tau^{-1/2}$. The expected stability from quantum projection noise (dashed-dotted green line) is \divqpn\,\% lower.  }
		\label{fig:LLI_results}
	\end{figure}

	The data acquired over a period of \timespan\,h, with an up-time of \uptime\,h, is shown in Fig.\,\ref{fig:LLI_results}\,a. 
	Data points are averaged in equidistant bins of \tbunch\,min. The measured population is decomposed in two parts $\pf=P(\kquad)+P(\klv)$, where $P(\kquad)$ gives rise to a constant offset of $P_0=0.39$ and $P(\klv)=\mathrm{d}\pf/\mathrm{d}\kappa\times \klv$ contains a potential LV signal. To extract $\klv$ from $\pf$, a high-pass filter is applied with a cutoff frequency of $\nu_c=5$\,$\mu$Hz, removing $P_0$ and slow variations ($\tau_{\mathrm{drift}}<2.5$\,days) caused by drifts in the E3 excitation probability, and the slope $\mathrm{d}\pf/\mathrm{d}\kappa\vert_{\kappa=\kappaVal}=-4.4(4)$ is calculated. More details on the data handling and the measurement sensitivity can be found in the Supplementary Information.\\
	In search of LV at Fourier frequencies of $\omegaSD$ and $2\omegaSD$, the data is fitted globally to the fit function
	\begin{equation}
		\label{eq:fit_func}
		\begin{aligned}
			\klv= &2\pi\times 5.1\times10^{15}\times[-3\sin(2\chi)c_{XZ}\cos(\omegaSD T)-3\sin(2\chi)c_{YZ}\sin(\omegaSD T)\\& 
			-\frac{3}{2}c_{X-Y}\sin^2(\chi)\cos(2\omegaSD T)-3c_{XY}\sin^2(\chi)\sin(2\omegaSD T)]\,,
		\end{aligned}
	\end{equation}
	from which the individual components of the $\cmn$ tensor in the SCCEF are extracted, where $c_{X-Y}=c_{XX}-c_{YY}$. The fit overlays the data in Fig.\,\ref{fig:LLI_results}\,a and the residuals from the fit are shown in Fig.\,\ref{fig:LLI_results}\,b, from which the reduced chi-square of $\chi^2=\chsqfit$ is extracted. The fitted values of the components of the $\cmn$ tensor are given in Table\,\ref{tab:results_cmunu}. For comparison, the values obtained in refs.~\citen{megidish_improved_2019,sanner_optical_2019} are also presented. The uncorrelated linear combination of the fit parameters, calculated by diagonalizing the covariance matrix, are given in Table\,\ref{tab:results_uncor_cmunu}. The spectral content of the data, shown in Fig.\,\ref{fig:LLI_results}\,c, is analyzed using a Lomb-Scargl periodogram \cite{glynn_detecting_2005}, which is specifically suited for spectral analysis of irregularly spaced data. The statistical significance level, i.e.~p-values \cite{glynn_detecting_2005}, of $50$\,$\%$ and $95$\,$\%$ are indicated by horizontal lines.\\ 
	The fit results show that the extracted values for $c_{X-Y}$, $c_{XZ}$ and $c_{YZ}$ are consistent with zero within a $1\sigma$ uncertainty. Only $c_{XY}$ shows a $1.1\sigma$ deviation from zero, but spectral analysis does not show a significant Fourier component at $2\omegaSD$. Therefore, we conclude that we do not find evidence of LV, in agreement with previous work \cite{megidish_improved_2019,sanner_optical_2019}. The stability of the data points is analyzed by calculating the Allan deviation, as shown in Fig.\,\ref{fig:LLI_results}\,d. The data averages down as $\sigma_{\kappa}=372(9)\,\mathrm{mrad\,s^{-1}}\,\tau^{-1/2}$, which is \divqpn\,\% higher than what is expected from quantum projection noise.\\
	\begin{table}[h]
		\centering
		\begin{tabular}{llll} 
			\hline& \\[-1.5ex]
			correlated LV & this work & $^{171}$Yb$^+$ limits \cite{sanner_optical_2019} & $^{40}$Ca$^+$ limits \cite{megidish_improved_2019} \\
			\multicolumn{4}{l}{parameters}\\
			\hline& \\[-1.5ex]
			$c_{X-Y}$ & $(-5.2\pm7.8)\times 10^{-21}$ & $(-0.5\pm 1.7)\times 10^{-20}$ & $(6.2\pm9.2)\times10^{-19}$\\ 
			$c_{XY}$ & $(4.4\pm3.9)\times 10^{-21}$ & $(-7.0\pm 8.1)\times 10^{-21}$ & $(2.4\pm4.8)\times10^{-19}$\\
			$c_{XZ}$ & $(-5.0\pm9.3)\times 10^{-21}$ & $(0.8 \pm 1.3)\times 10^{-20}$ & $(0.8\pm2.1)\times10^{-19}$\\
			$c_{YZ}$ & $(6.3\pm8.9)\times 10^{-21}$ & $(1.0 \pm 1.3)\times 10^{-20}$ & $(-3.1\pm2.2)\times10^{-19}$\\
			\hline& \\[-1.5ex]
		\end{tabular}
		\caption{The symmetry breaking tensor components. The correlated components of the $\cmn$ tensor extracted from the fit shown in Fig.\,\ref{fig:LLI_results}\,a, compared to existing limits from refs.~\cite{sanner_optical_2019} and \cite{megidish_improved_2019}. All uncertainties represent a $1\sigma$ interval.}
		\label{tab:results_cmunu}
	\end{table}
	\begin{table}[h]
		\centering
		\begin{tabular}{ll} 
			\hline& \\[-1.5ex]
			uncorrelated linear combinations of parameters & this work \\ [1ex]
			\hline& \\[-1.5ex]
			$0.70c_{XZ}+0.26c_{YZ}+0.51c_{X-Y}+0.42c_{XY}$& $(0.0\pm1.0)\times10^{-20}$\\
			$0.24c_{XZ}-0.61c_{YZ}+0.46c_{X-Y}-0.59c_{XY}$& $(-13.8\pm9.6)\times10^{-21}$\\
			$-0.47c_{XZ}-0.50c_{YZ}+0.38c_{X-Y}+0.62c_{XY}$& $(3.0\pm7.2)\times10^{-21}$\\
			$-0.48c_{XZ}+0.56c_{YZ}+0.61c_{X-Y}-0.29c_{XY}$& $(-0.4\pm3.5)\times10^{-21}$\\
			\hline& \\[-1.5ex]
		\end{tabular}
		\caption{The uncorrelated combination of symmetry breaking tensor components. The covariance matrix from the fit shown in Fig.\,\ref{fig:LLI_results}\,a is diagonalized to obtain the linear combination of parameters of the $\cmn$ tensor. All uncertainties represent a $1\sigma$ interval.}
		\label{tab:results_uncor_cmunu}
	\end{table}
	\subsection{Discussion}
	The presented results set the most stringent bounds on this type of LV in the combined electron-photon sector. With the presented method, the resolution of the previous most sensitive measurement \cite{sanner_optical_2019} is reached 9 times faster. We improve on the state-of-the-art by a factor of $2.2$ and constrain all the coefficients of the $\cmn$ tensor now at the $10^{-21}$ level. Due to the experimental geometry, a higher sensitivity is reached for signals that oscillate at $2\omegaSD$ than those that oscillate at $\omegaSD$, see the Supplementary Information. Therefore, the tightest constraint of $3.9\times10^{-21}$ is achieved on $c_{XY}$. \\ 
	In this work, coefficients of the first and second harmonic order of the sidereal day modulation frequency were considered. However, due to the high total angular momentum of the $^2F_{7/2}$ state, the applied method is sensitive to LV at harmonics of up to sixth order \cite{kostelecky_lorentz_2018,vargas_overview_2019}. Therefore, in combination with improved many-body calculations, experimental constraints can be translated into bounds on a larger number of coefficients in the future \cite{vargas_overview_2019}.\\
	The method demonstrated in this work is technically less demanding and more robust than alternative methods requiring simultaneous operation of independent optical clocks or high-fidelity entanglement gates \cite{sanner_optical_2019,megidish_improved_2019}. It is applicable in a wide variety of systems, e.g., highly-charged ions or ultra-cold atoms \cite{shaniv_new_2018} and it is capable of scaling to multiple $N_{\mathrm{ion}}$ ions in linear ion Coulomb crystals for a further enhancement of the sensitivity. Up-scaling with an entanglement-based techniques \cite{megidish_improved_2019} is technically demanding when considering the $^2F_{7/2}$ state in Yb$^+$ due to the low-fidelity ($p$) single ion gate on the E3 transition, which further deteriorates, according to $p^{N_{\mathrm{ion}}}$, for larger ion numbers. In contrast, the implemented composite rf pulse sequence is highly robust against errors originating from spatial and temporal fluctuations of both the ambient field and the rf field. Therefore, the coherence time is not expected to significantly decrease for larger ion numbers and a $\sqrt{N_{\mathrm{ion}}}$ higher sensitivity is expected in such a system. Moreover, the long radiative lifetime of the $^2F_{7/2}$ state \cite{lange_lifetime_2021} does not significantly limit the coherence time and, with several technical improvements, longer interrogation times could be reached. Note that for efficient population transfer via the E3 transition, advanced cooling techniques, such as EIT \cite{morigi_ground_2000} or Sisyphus cooling \cite{ejtemaee_3d_2017,joshi_polarization_2020}, might be advantageous in larger ion crystals. With the presented method, the boundaries of Lorentz symmetry tests can be pushed to the 10$^{-22}$ level with a string of $10$ Yb$^+$ ions in the future.
	\clearpage

	\section*{Methods}
	\label{sec:methods}
	
	\subsection*{Experimental details}
	A single $^{172}$Yb$^{+}$ ion is trapped in a segmented rf Paul-trap \cite{pyka_high-precision_2014,keller_probing_2019}. The radial confinement is set with an rf electric field supplied by a resonant circuit at a frequency of $\Omega_{\mathrm{rf}}/2\pi= 24.38$\,MHz, while the axial confinement is set by dc voltages supplied to the trapping segment and the neighboring segments. With the applied confinement, the secular frequencies are $\omega(\mathrm{rad1},\mathrm{rad2},\mathrm{ax})/2 \pi =(775,510,266)$\,kHz. Micromotion is measured on a daily basis with the photon correlation technique \cite{keller_precise_2015} and compensated in three directions to typically $E_{\mathrm{rf}}<100$\,V/m. The quantization field of $\mathbf{B}=221$\,$\mu$T lies in the horizontal plane under an angle of $25^{\circ}$ to the trap axis. The $\mathbf{B}$-field is measured with a sensor near the vacuum chamber and active feedback is applied in three orthogonal directions via current modulation of six auxiliary coils.\\ 
	The ion is cooled to approximately 0.5\,mK, close to the Doppler limit, on the dipole allowed $^2S_{1/2}\rightarrow^2P_{1/2}$ transition assisted by a repumper laser near $935$\,nm. Fluorescence from the decay of the $^2P_{1/2}$ state is collected by a lens of $\mathrm{N/A} = 0.27$ and imaged onto the electron-multiplying charge-coupled device (EMCCD) camera \cite{pyka_high-precision_2013}. This enables state detection via the electron shelving technique. The ion can be prepared in either the $\ket{^2S_{1/2},m_j=-1/2}$ or the $\ket{^2S_{1/2},m_j=+1/2}$ state using circular polarized beam near $370$\,nm pointing along the direction of the quantization axis. The 467\,nm laser for excitation on the E3 transition lies in the radial plane and its beam waist is $26(3)\times38(3)$\,$\mu$m at the ion. The power is stabilized to $6.0(2)$\,mW, at which a Rabi frequency of $\Omega_{\mathrm{E3}}/2\pi=10$\,Hz is reached. The light is frequency-shifted and pulsed using acousto-optic modulators.\\ 
	The resonant rf coil, consisting of 27 turns wound at a diameter of $4.5$\,cm, is placed $5.5$\,cm above the ion. The resonance frequency of the coil is $\omega_{\mathrm{res}}/2\pi=3.5147(7)$\,MHz and it is driven by a signal derived from a direct digital synthesizer (DDS), referenced to a stable $100$\,MHz signal from a hydrogen maser. At $221$\,$\mu$T, the resonance frequency between the $m_j$ levels in the $^2F_{7/2}$ state is $\omega_{\mathrm{rf}}/2\pi=3.52$\,MHz, close to the resonance frequency of the coil. The achieved multi-level Rabi frequency is $\Omega_{F}/2\pi=33$\,kHz. The ambient magnetic field is monitored throughout the measurement campaign via data acquired in the E3 servo sequence. Drifts are observed at the level of $100$\,nT, corresponding to a detuning $\Delta \omega_{\mathrm{rf}}/2\pi=\rfFreqDrift$\,kHz. The frequency supplied to the coils is actively adjusted to remain in resonance with the frequency given by the linear Zeeman splitting. For this purpose, the resonance frequency as a function of magnetic field was calibrated to be $\omega_{\mathrm{rf}}(B)/2\pi=\left[1.581(1.6) B+0.0162(3)\right]$ in MHz. For further details on the experimental set-up, see refs.~\citen{pyka_high-precision_2013,pyka_high-precision_2014,fuerst_coherent_2020,kalincev_motional_2021}.

	\begin{itemize}
		\item \textbf{Data availability}
		Reasonable requests for source data should be addressed to the corresponding author.
		\item \textbf{Code availability} 
		Reasonable requests for computer code should be addressed to the corresponding author.
		\item \textbf{Acknowledgments} We thank Melina Filzinger, Richard Lange, Burkhard Lipphardt, Nils Huntemann and Andr\'{e} Kulosa for experimental support. We thank Ralf Lehnert, Arnaldo Vargas, Nils Huntemann and Ekkehard Peik for helpful discussions and Ravid Shaniv for motivating this work. We thank Ralf Lehnert for carefully reading the manuscript.\\
		L.S.D. acknowledges support from the Alexander von Humboldt foundation. This project has been funded by the Deutsche Forschungsgemeinschaft (DFG, German Research Foundation) under Germany’s Excellence Strategy – EXC-2123 QuantumFrontiers –390837967 (RU B06) and through Grant No. CRC 1227 (DQ-mat, project B03). This work has been supported by the Max-Planck-RIKEN-PTB-Center
		for Time, Constants and Fundamental Symmetries. 
		\item \textbf{Author contributions}
		Conception of the experiment and development of methods: L.S.D., C-H.Y., H.A.F., K.C.G., T.E.M. Design and construction of experimental apparatus: L.S.D., C-H.Y., H.A.F., T.E.M. Data acquisition and analysis: L.S.D., C-H.Y. Preparation and discussion of the manuscript: L.S.D., C-H.Y., H.A.F., K.C.G., T.E.M.
		\item \textbf{Competing interests}
		The authors declare no competing interests.
	\end{itemize}
	\clearpage
	
	\setcounter{figure}{0}
	\setcounter{table}{0}
	\makeatletter 
	\renewcommand{\thefigure}{S\@arabic\c@figure}
	\renewcommand{\thetable}{S\@arabic\c@table}
	\makeatother

	\section*{Supplementary Note 1}
	\label{sec:to_SCCEF_frame}
	The elements of the $\cmunu$ tensors are dependent of the local laboratory frame. In order to compare our results with those from other experiments, we transform the components of the local $c_{mn}$ tensor to the sun-centered, celestial-equatorial frame (SCCEF). We follow a similar derivation as given in ref.~\citen{sanner_optical_2019}.\\
	The relation between the components of $c_{mn}$ in the lab frame and $\cmn$ in the SSCEF is given by
	\begin{equation}
		c_{mn}=\Lambda^{M}_{m}\Lambda^{N}_{n}c_{MN}\,,
	\end{equation} 
	where $\Lambda$ is the Lorentz transformation matrix, which consists of rotations and boosts in the lab frame relative to the SCCEF. Our lab frame has its origin at a colatitude of $\chi=37.7^{\circ}$ and a longitude of $\lambda=10.5^{\circ}$  (PTB Braunschweig, Germany). In the local coordinate system, $\hat{x}$ points towards the east, $\hat{y}$ points towards the north and $\hat{z}$ points upward. The rotation matrix that maps the SCCEF coordinate frame to that of the lab is given by
	\begin{equation}
		\label{eq:rotmat}
		\mathbf{R} = 
		\begin{pmatrix}
			-\sin(\omegaSD T) & \cos(\omegaSD T) & 0 \\
			-\cos(\chi)\cos(\omegaSD T) & -\cos(\chi)\sin(\omegaSD T) &  \sin(\chi) \\
			\sin(\chi)\cos(\omegaSD T) & \sin(\chi)\cos(\omegaSD T) &  \cos(\chi)
		\end{pmatrix}\,,
	\end{equation}
	where $\omegaSD/2\pi=1/23.934$\,h is the angular frequency of a sidereal day. Our quantization magnetic field $B$ lies in the $xy$ plane and points $20^{\circ}$ south of east. Similar as in ref.~\citen{sanner_optical_2019}, we calculate a virtual location on Earth's surface, where $B$ points vertically upward by only changing the origin of the coordinate system, not affecting transformation formulas between the lab frame and the SCCEF frame. This yields $\chi_n=102.1^{\circ}$ and $\lambda_n=84.4^{\circ}$. Substituting $\chi=\chi_n$ and $\lambda=\lambda_n$ values in equation~(\ref{eq:rotmat}) and setting $T=0$ at the moment the $z$ axis of the new location points towards the Sun on the day of the Vernal equinox, in this case 03:59:24 UTC, March 20, 2021 yields the proper transformation equations.\\
	The boost of the experimental frame as seen from the SCCEF is given by
	\begin{equation}
		\pmb{\beta} = 
		\begin{pmatrix}
			\beta \sin(\Omega_{\odot} T) \\
			-\beta\cos(\eta)\cos(\Omega_{\odot} T)\\
			-\beta\sin(\eta)\cos(\Omega_{\odot} T)
		\end{pmatrix}\,,
	\end{equation}
	where $\Omega_{\odot}/2\pi=1/(365.256 \times 24 \mathrm{h})$ is the angular frequency of a sidereal year, $\beta=1\times10^{-4}$ is the magnitude of the boost from the orbital velocity and $\eta=23.4^{\circ}$ is the angle between the ecliptic plane and the Earth's equatorial plane. Here the boost from the Earth's rotation has been neglected as it is two orders of magnitude smaller ($\beta_L=1.5\times10^{-6}$).
	\begin{table}[h]
		\centering
		\begin{tabular}{c c c} 
			\hline& \\[-1.5ex]
			$\omega_j$ & $C_j$ & $S_j$ \\ [1ex] 
			\hline& \\[-1.5ex]
			$\omegaSD$ & $-3\sin(2\chi)c_{XZ}+2c_{TY}\beta_L$ & $-3\sin(2\chi)c_{YZ}-2c_{TX}\beta_L$ \\ 
			$2\omegaSD$ & $-\frac{3}{2}(c_{XX}-c_{YY})\sin^2(\chi)$ & $-3c_{XY}\sin^2(\chi)$ \\
			$\Omega_{\odot}$ & $-\frac{1}{2}\beta(3\cos(2\chi)+1)(c_{TY}\cos(\eta)-2c_{TZ}\sin(\eta))$ & $\frac{1}{2}\beta c_{TX}(3\cos(2\chi)+1)$ \\
			$2\Omega_{\odot}$ & $0$ & $0$ \\
			$\Omega_{\odot}-\omegaSD$ & $\frac{3}{2}\beta c_{TX}\sin(\eta)\sin(2\chi)$ & $-\frac{3}{2}\beta \sin(2\chi)[c_{TY}\sin(\eta)+c_{TZ}(1+\cos(\eta))]$ \\
			$\Omega_{\odot}+\omegaSD$ & $\frac{3}{2}\beta c_{TX}\sin(\eta)\sin(2\chi)$ & $-\frac{3}{2}\beta \sin(2\chi)[c_{TZ}(1-\cos(\eta))-c_{TY}\sin(\eta)]$ \\
			$2\Omega_{\odot}-\omegaSD$ & $0$ & $0$ \\
			$2\Omega_{\odot}+\omegaSD$ & $0$ & $0$ \\
			$\Omega_{\odot}-2\omegaSD$ & $-3\beta c_{TY}\cos^2(\eta/2)\sin^2(\chi)$ & $-3\beta c_{TX}\cos^2(\eta/2)\sin^2(\chi)$ \\
			$\Omega_{\odot}+2\omegaSD$ & $3\beta c_{TY}\sin^2(\eta/2)\sin^2(\chi)$ & $-3\beta c_{TX}\sin^2(\eta/2)\sin^2(\chi)$ \\
			$2\Omega_{\odot}-2\omegaSD$ & $0$ & $0$ \\
			$2\Omega_{\odot}+2\omegaSD$ & $0$ & $0$ \\ [1ex] 
			\hline& \\[-1.5ex]
		\end{tabular}
		\caption{Contributions of $\cmn$ to $C_{0}^{(2)}$. The angular frequencies and the corresponding amplitudes contributing to $C_{0}^{(2)}$ in the SCCEF as a function of the colatitude $\chi$ and the angle between the ecliptic plane and the Earth's equatorial plane $\eta$.}
		\label{tab:amplitudesCjSj}
	\end{table}
	Using the rotations and boosts, the Lorentz transformation that maps $\cmunu$ from the SCCEF to the lab frame is given by 
	\begin{equation}
		\label{eq:LTM}
		\Lambda = 
		\begin{pmatrix}
			1 & -\beta^1 & -\beta^2 & -\beta^3 \\
			-(R\cdot\beta)^1 & R^{11} & R^{12} & R^{13} \\
			-(R\cdot\beta)^2 & R^{21} & R^{22} & R^{23} \\
			-(R\cdot\beta)^3 & R^{31} & R^{32} & R^{33} 
		\end{pmatrix}\,,
	\end{equation}	
	Using equation~(\ref{eq:LTM}), the parameter $C_{0}^{(2)}$ can now be expressed in terms of components of $\cmn$ in the SCCEF using the Lorentz transformation. It is given by
	\begin{equation}
		\Clv=A_0+\sum_{j}[C_j\cos(\omega_j T)+S_j\sin(\omega_j T)],
	\end{equation}
	where $A_0$ is a constant offset, $\omega_j$ contains all linear combinations of $\omegaSD$ and $\Omega_{\odot}$, and $C_j$ and $S_j$ are the respective amplitudes as given in Tab.~\ref{tab:amplitudesCjSj}.\\
	With the high-pass filter applied to the data, we are sensitive only to signals that oscillate at frequencies larger than $\nu_c=5$ $\mu$Hz. Therefore the Lorentz violating signal is given by 
	\begin{equation}
		\label{eq:C02SCCEF}
		\begin{aligned}
			\Clv&=-3\sin(2\chi)c_{XZ}\cos(\omegaSD T)-3\sin(2\chi)c_{YZ}\sin(\omegaSD T) \\
			& -\frac{3}{2}(c_{XX}-c_{YY})\sin^2(\chi)\cos(2\omegaSD T)-3c_{XY}\sin^2(\chi)\sin(2\omegaSD T).
		\end{aligned}
	\end{equation}
	Combining equation~(\ref{eq:C02SCCEF}) with the sensitivity of $\klv$ to $C_0^{(2)}$ \cite{shaniv_new_2018} yields
	\begin{equation}
		\label{eq:kappaSCCEF}
		\begin{aligned}
			\kappa_{LV}&=2\pi\times 5.1\times10^{15}\times[-3\sin(2\chi)c_{XZ}\cos(\omegaSD T)-3\sin(2\chi)c_{YZ}\sin(\omegaSD T) \\
			& -\frac{3}{2}(c_{XX}-c_{YY})\sin^2(\chi)\cos(2\omegaSD T)-3c_{XY}\sin^2(\chi)\sin(2\omegaSD T)]\,.
		\end{aligned}
	\end{equation}
	
	The stability of $\klv$ was measured to be $\sigma_{\kappa}=372(9)$\,mrad\,s$^{-1}$$\tau^{-1/2}$, from which we can extract the stability of $\Clv$ to be $\sigma_{\Clv}=1.16\times10^{-17}$\,$\tau^{-1/2}$.
	
	\section*{Supplementary Note 2}
	\label{sec:measurement_sensitivity}
	The quadratic term in the free Hamiltonian for state $\ket{J,m_j}$ interacting with a magnetic field $\mathbf{B}=B_z\hat{z}$ is given by $\mathcal{H}_{\mathrm{quad}}=\kappa J_z^2$. It scales with $\kappa=\kquad+\klv$, where the first term is given by the quadrupole shift and the second term is given by a potential LV. The quadrupole shift can be calculated using \cite{roos_precision_2006}
	\begin{equation}
		\label{eq:quad_shift}
		\Delta \nu_{\mathrm{quad}} =\frac{1}{4}\frac{J(J+1)-3m_j^2}{J(2J-1)}\frac{1}{h} \Theta\left(^2F_{7/2}\right) \frac{\mathrm{d}E}{\mathrm{d}z}[3\cos^2(\beta)-1]\,,
	\end{equation}
	The quadrupole moment of the $^2F_{7/2}$ state is given by $\Theta\left(^2F_{7/2}\right)=-0.0297(5) e a_0^2$ \cite{lange_coherent_2020}, where $e$ is the electron charge and $a_0$ is the Bohr radius. The electric field gradient is given by $\mathrm{d}E/\mathrm{d}z=-m_{\mathrm{ion}}\omega_z^2/q$ for a single trapped ion with mass $m_{\mathrm{ion}}$ and charge $q$ at the equilibrium position of the trap.
	\begin{figure}[h]
		\centering	
		\includegraphics[width=1.0\textwidth]{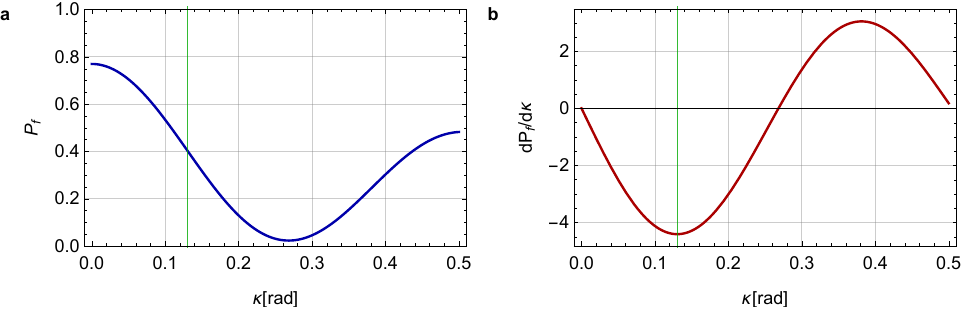}
		\caption{The sensitivity of the measured quantity $\pf$ to variations in $\kappa$. \textbf{a} The calculated final population $\pf$ as a function of $\kappa$ at $T=1.15$ s, where the experimentally achieved contrast of 0.77(6) was taken into account. \textbf{b} The derivative $\mathrm{d}\pf/\mathrm{d}\kappa$ as a function of $\kappa$ at $T=1.15$ s. The measurement sensitivity at $\kquad=0.13$ (green vertical line in \textbf{a} and \textbf{b}) is given by $\mathrm{d}\pf/\mathrm{d}\kappa\vert_{\kappa=0.13}=-4.4$. }
		\label{fig:RamseyDTKappa}
	\end{figure}
	The angle between the quantization axis and the principle axis of the trap is in our case $\beta=25^{\circ}$. The corresponding value for $\kquad$ is given by
	\begin{equation}
		\frac{\kquad}{2\pi}=-\frac{3}{4J(2J-1)}\frac{1}{h}\Theta\left(^2F_{7/2}\right)\frac{\mathrm{d}E}{\mathrm{d}z}[3\cos^2(\beta)-1]\,.
	\end{equation}
	At typical values of the axial secular frequency in our trap $\omega_{\mathrm{ax}}/2\pi= (200-290)\,\mathrm{kHz}$, the quadrupole shift is $\nu_{\mathrm{quad}}=60-125$\,mHz, corresponding to $\kquad = 0.075-0.150$\,rad/s.\\
	The fraction $\pf$ retrieved back into the $m_j=\pm1/2$ state at the end of the rf sequence is dependent on the acquired phase $\kappa T_{\mathrm{D}}$, where $T_{\mathrm{D}}$ is the Ramsey dark time. At $\kappa T_{\mathrm{D}} = 0.15$ rad, the slope $\vert\mathrm{d}\pf/\mathrm{d}\kappa\vert$ is maximum and the highest measurement sensitivity is reached. In the experiment, $\kappa=\kappaVal$ for which an optimum Ramsey dark-time of $T_{\mathrm{D}}=\darktime$\,s is found. Using the average achieved contrast of $(P_{\mathrm{f,max}}-P_{\mathrm{f,min}})/(P_{\mathrm{f,max}}+P_{\mathrm{f,min}})=0.77$, $\pf$ and $\mathrm{d}\pf/\mathrm{d}\kappa$ are calculated as a function of $\kappa$, as shown in Fig.\,\ref{fig:RamseyDTKappa}\,a and b, respectively. At $\kquad=\kappaVal$, the sensitivity of $\pf$ to variations of $\kappa$ is calculated to be $\mathrm{d}\pf/\mathrm{d}\kappa\vert_{\kappa=\kappaVal}=-4.4(4)$. The uncertainty on $\klv$ stemming from $\Delta (\mathrm{d}\pf/\mathrm{d}\kappa\vert_{\kappa=\kappaVal})=0.4$ is added to $\Delta \klv$ in quadrature.
	
	\section*{Supplementary Note 3}
	\label{sec:data_analysis}
\begin{figure}[h]
		\centering	
		\includegraphics[width=0.5\textwidth]{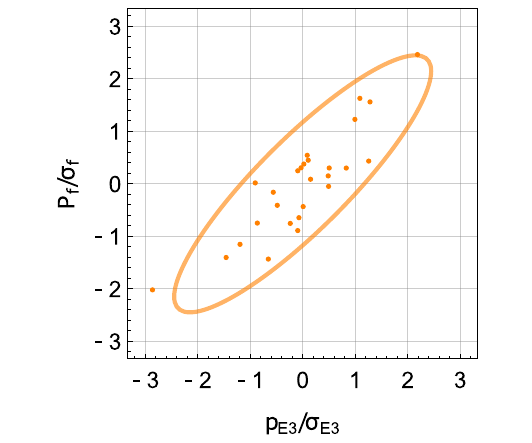}
		\caption{Correlation plot. The correlation between the E3 excitation probability $p_{\mathrm{E3}}$ and the measured quantity $\pf$ averaged over a time span of about one day is plotted. The data points show $p_{\mathrm{E3}}/\sigma_{\mathrm{E3}}$ and $\pf/\sigma_{\mathrm{f}}$ after subtracting the offset of $22.6$ and $9.2$, respectively. Pearson's correlation is calculated to be $0.90$. The ellipse indicates the $95\%$ confidence interval. }
		\label{fig:correlations}
	\end{figure}

	The E3 servo sequence is based on four measured populations at half the linewidth of the two opposite Zeeman transitions $\ket{^2S_{1/2},m_j=\pm1/2}\rightarrow\ket{^2F_{7/2},m_j=\pm1/2}$. From the average value of the four measured data points the excitation probability of the E3 transition ($p_{\mathrm{E3}}$) can be calculated. The  E3 servo sequence is repeated every 50 data points throughout the measurement campaign, allowing us to monitor $p_{\mathrm{E3}}$ during this time.\\ 
	Slow drifts of $p_{\mathrm{E3}}$ are observed on timescales of $\tau<2.5$\,days, corresponding to Fourier frequencies of $\omega/2\pi< 5$\,$\mu$Hz, due to changes in, e.g.,~beam pointing and ambient noise. The quantity $\pf$ is detected via de-excition from the $^2F_{7/2}$ on the E3 transition and it is, therefore, highly correlated with $p_{\mathrm{E3}}$. To quantify the correlation, data points are averaged over a time span of about one day and Pearson's correlation factor is calculated to be $0.9$. The number of measurements per averaged data point are not equal for $p_{\mathrm{E3}}$ and $\pf$. Therefore, the standard deviation of the two data sets are significantly different. For visualization purposes, the measured quantities are scaled by their respective standard deviation and plotted together with the $95\%$ confidence interval, see Fig.\,\ref{fig:correlations}. Note that Pearson's correlation factor differs from 1, because the $p_{\mathrm{E3}}$ and $\pf$ are not measured at exactly the same time, but rather in an alternating fashion.
	\\
	The measured data points $\pf$ are corrected for slow drifts of $p_{\mathrm{E3}}$. Residual slow variations that are not clearly connected to $p_{\mathrm{E3}}$ are observed in the data at $\omega/2\pi=1.65$\,$\mu$Hz, related to a fluctuation on the timescale of a week. This Fourier component is removed from the data with a high-pass filter using a Hamming window with a cut-off frequency of $\nu_{c}=5$\,$\mu$Hz. Bounds on LV at frequencies $\omegaSD$ and $2\omegaSD$ are extracted from the filtered data using a fit to equation (\ref{eq:kappaSCCEF}).\\
	To investigate the possible influence from the filter on the result, data points are simulated with Fourier components at $\omega/2\pi=1.5$\,$\mu$Hz, $\omega=\omegaSD$ and $\omega=2\omegaSD$, mimicking the observed slow drifts and a hypothetical LV signal, respectively. Exactly the same time stamps are used for the simulated data as for the experiment. The high-pass filter is applied to the simulated data with different values of $\nu_{c}$, after which it is fitted to extract the amplitudes at $\omegaSD$ and $2\omegaSD$. The retrieved amplitudes do not significantly differ from the simulated amplitudes for cut-off frequencies between $2<\nu_{c} <10$\,$\mu$Hz. To validate this, the actual experimental data is also filtered with different values of $\nu_{c}$ in the range of $2<\nu_{c} <10$\,$\mu$Hz. The extracted amplitudes at $\omegaSD$ and $2\omegaSD$ from the fit to equation (\ref{eq:kappaSCCEF}) do not show a significant deviation for different values of $\nu_{c}$ in this range.

\end{document}